\documentstyle[12pt]{article}
\def\PRL #1 #2 #3{{\sl Phys. Rev. Lett.} {\bf#1} (#2) #3}
\def\NPB #1 #2 #3{{\sl Nucl. Phys.} {\bf B#1} (#2) #3}
\def\NPBFS #1 #2 #3 #4{{\sl Nucl. Phys.} {\bf B#2} [FS#1] (#3) #4}
\def\CMP #1 #2 #3{{\sl Commun. Math. Phys.} {\bf #1} (#2) #3}
\def\PRD #1 #2 #3{{\sl Phys. Rev.} {\bf D#1} (#2) #3}
\def\PLA #1 #2 #3{{\sl Phys. Lett.} {\bf #1A} (#2) #3}
\def\PLB #1 #2 #3{{\sl Phys. Lett.} {\bf #1B} (#2) #3}
\def\JMP #1 #2 #3{{\sl J. Math. Phys.} {\bf #1} (#2) #3}
\def\PTP #1 #2 #3{{\sl Prog. Theor. Phys.} {\bf #1} (#2) #3}
\def\SPTP #1 #2 #3{{\sl Suppl. Prog. Theor. Phys.} {\bf #1} (#2) #3}
\def\AoP #1 #2 #3{{\sl Ann. of Phys.} {\bf #1} (#2) #3}
\def\PNAS #1 #2 #3{{\sl Proc. Natl. Acad. Sci. USA} {\bf #1} (#2) #3}
\def\RMP #1 #2 #3{{\sl Rev. Mod. Phys.} {\bf #1} (#2) #3}
\def\PR #1 #2 #3{{\sl Phys. Reports} {\bf #1} (#2) #3}
\def\AoM #1 #2 #3{{\sl Ann. of Math.} {\bf #1} (#2) #3}
\def\UMN #1 #2 #3{{\sl Usp. Mat. Nauk} {\bf #1} (#2) #3}
\def\FAP #1 #2 #3{{\sl Funkt. Anal. Prilozheniya} {\bf #1} (#2) #3}
\def\FAaIA #1 #2 #3{{\sl Functional Analysis and Its Application} {\bf
#1} (#2) #3}
\def\BAMS #1 #2 #3{{\sl Bull. Am. Math. Soc.} {\bf #1} (#2)
#3} \def\TAMS #1 #2 #3{{\sl Trans. Am. Math. Soc.} {\bf #1} (#2) #3}
\def\InvM #1 #2 #3{{\sl Invent. Math.} {\bf #1} (#2) #3}
\def\LMP #1 #2 #3{{\sl Letters in Math. Phys.} {\bf #1} (#2) #3}
\def\IJMPA #1 #2 #3{{\sl Int. J. Mod. Phys.} {\bf A#1} (#2) #3}
\def\AdM #1 #2 #3{{\sl Advances in Math.} {\bf #1} (#2) #3}
\def\RMaP #1 #2 #3{{\sl Reports on Math. Phys.} {\bf #1} (#2) #3}
\def\IJM #1 #2 #3{{\sl Ill. J. Math.} {\bf #1} (#2) #3}
\def\APP #1 #2 #3{{\sl Acta Phys. Polon.} {\bf #1} (#2) #3}
\def\TMP #1 #2 #3{{\sl Theor. Mat. Phys.} {\bf #1} (#2) #3}
\def\JPA #1 #2 #3{{\sl J. Physics} {\bf A#1} (#2) #3}
\def\JSM #1 #2 #3{{\sl J. Soviet Math.} {\bf #1} (#2) #3}
\def\MPLA #1 #2 #3{{\sl Mod. Phys. Lett.} {\bf A#1} (#2) #3}
\def\JETP #1 #2 #3{{\sl Sov. Phys. JETP} {\bf #1} (#2) #3}
\def\JETPL #1 #2 #3{{\sl  Sov. Phys. JETP Lett.} {\bf #1} (#2) #3}
\def\PHSA #1 #2 #3{{\sl Physica} {\bf A#1} (#2) #3}
\def\CQG #1 #2 #3{{\sl Class. Quantum Grav.} {\bf #1} (#2) #3}
\def\SJNP #1 #2 #3{{\sl Sov. J. Nucl. Phys. (Yadern.Fiz.)} {\bf #1} (#2) #3}
\def\a{\alpha}\def\g{\gamma}
\def\E{\varepsilon}
\def\l{\lambda}\def\s{\sigma}
\def\G{\Gamma}
\def\p{\partial}\def\t{\tau}\def\vp{\varphi}

\begin{document}
\unitlength=1mm

\vspace{20mm}

\title{
Tension as a perturbative parameter in
non--linear \\ string equations in curved
space--time}
\author{\sl A.A.Zheltukhin,\\ NSC
Kharkov Institute of Physics and Technology, \\ 310108, Kharkov,
Ukraine} \date{} \maketitle

\begin{center}
\parbox{140mm}{\small A perturbation theory with respect to the
tension parameter $\g/\a^\prime$ for the non--linear \\ equations of
string, moving in curved space--time, is considered. Obtained are \\
linearized motion equations for the functions of the $n-$th degree of
approximation ($n=0,1,2$).} \end{center}

\bigskip
\bigskip
\bigskip

The study of string dynamics in curved space--time has evoked considerable
interest [1--3]. The investigation of this problem is hampered by
a non--linear character of the string equations exactly solvable for a
restricted class of special metrics [1--8]. Therefore in [2,3,7] it was
proposed to study the approximate solutions of these equations  using a
perturbative theory. A rigorous realization of this program implies the
presence of a small parameter in the string  equations, and the search for
their solutions in the form of a series expansion in terms of this small
parameter.

         Here it will be shown that the string tension may play the role of a
small parameter in the string motion equations. We will prove that the
solution of the string equations and constraints can be sought in the form
of a series expansion in terms of the tension parameter. The functions of
the zero and subsequent approximations depend on the singular world--sheet
variable $\xi = \s /\E $, where $\E $ is a small dimensionless parameter
proportional to the string tension.

      To investigate the described problem it is convenient to use the
reparametrization invariant representation [9] for the bosonic string action

\begin{equation}\label{1}
S=S_0+S_1 = \int d\t d\s \left[ {det (\partial_\mu x^M G_{MN}\p_\nu
x^N)\over E(\t ,\s)} - {1\over(\a^\prime)^2} E (\t ,\s)\right]~~,
\end{equation}
 where $G_{MN}(x) $ is the metric tensor  of $D$ -- dimensional target
 space--time, $E(\t ,\s) $ is an auxiliary field of the
 two--dimensional world sheet density. The parameter $T\equiv 1/\a^\prime
 $ in the world sheet $T^2E$ (1) has the physical sense as the string
 tension. The latter fact becomes evident if $E(\t ,\s) $ is eliminated from
the action (1) using the motion equation

 \begin{equation}\label{2} E= \a^\prime \sqrt{-det g_{\mu\nu}} ,
 ~~~ g_{\mu\nu} = \partial_\mu x^MG_{MN}\partial_\nu x^N , \end{equation}
which connects $E$ with the induced world sheet metric $g_{\mu\nu}(\t
,\s)$.  After eliminating $E$ the action (1) transforms into the well--known
Nambu--Goto representation. The representation (1) is suitable because it
contains the tension parameter $T$ as an multiplicative factor in addition
to the kinetic term $ S_0$ (1) which coincides with the total action for
null string [8--10]. In the limiting case $T\to 0$ (or $\a^\prime \to \infty
$) the additive term $S_1$ in Eq.(1) can be considered as a perturbation
term for the null string action $S_0$.

The introduction of the momentum  ${\cal P}_M$ canonically conjugate to
the string coordinates $x^M$ \begin{equation}\label{3} {\cal
P}_M = -{2\over E} G_{MN}[(x^{\prime L}x^{\prime}_L)\dot{x}^N -
(\dot{x}^Lx^{\prime}_L)x^{\prime N} ] \end{equation}
allows to present the Hamiltonian density
 ${\cal H}$ for the action   $S$ (1) in the form
 \begin{equation}\label{4}
 {\cal H} = -{E\over (x^{\prime L}x^{\prime}_L)} [{1\over 4}({\cal P}^M
 {\cal P}_M) + T^2(x^{\prime M}x^{\prime}_M)] + \l (x^{\prime M} {\cal
 P}_M), \end{equation}
 where $\l $ is the Lagrange multiplier for the constraint  $ ({\cal
 P}_M x^{\prime M}) =0 $.  As follows from Eq. (4), the fixation of the
 reparametrization gauge by the conditions for
  $E$ and $\l $
  \begin{equation}\label{5} E = -\g
 (x^{\prime L}x^{\prime}_L), ~~~~~~~~~~~~~ \l =0 ~~, \end{equation}
elucidates the physical sense of the world sheet "cosmological" term $S_1$
as the potential energy of string. Here $\g $ is a physical constant with the
dimension $L^2$ (if $\hbar =1=c$), characterizing  the energetic scale of
the universe;for example, we can put $ \g ={ 1\over M^2}$, where $M$ is the
mass of the considered universe. This potential term describes the general
covariant contribution of the potential energy of elastic string deformation
depending on the metric $G_{MN}(x)$. In the limiting case $T=0$ the
Hamiltonian density (4) is transformed into the density ${\cal H}_0 = \g /4
({\cal P}_M{\cal P}^M)$ corresponding to null string and generating a
continious mass spectrum [10].  The introduction of the oscillator addition
into  ${\cal H}_0$ leads to the appearance of oscillation regime, and to the
transformation of the continious spectrum of null string into the well-known
discrete spectrum of masses of the Nambu--Goto string.
    In view of the above mentioned interpretation of  ${\cal H}_0$ in the
gauge (5),i.e.
 \begin{equation}\label{6} E = -\g(x^{\prime L}
 x^\prime _L), \end{equation}

\begin{flushright}
\parbox{90mm}{
$(\dot{x}^L\cdot x^\prime _L)=0,$\hfill (6$^\prime$)}
\end{flushright}
it is clear that the non--linear motion equations derived from
 $S$ (1) are to be analyzed in this   gauge. In the gauge (6)
the equation (2) and the world--sheet metric  $g_{\mu\nu}$  are rewritten
in the form
 \begin{equation}\label{7} (\dot{x}^M\cdot
\dot{x}_M)+\left({\g\over \a^\prime}\right)^2 (x^{\prime M} x^\prime
_M)=0~,~~ g_{\mu\nu} = E diag [\g/(\a^\prime)^2, -1/\g] \end{equation}

As follows from Eq.(7), in the limiting case $T\to 0$ the world sheet
metric $ g_{\mu\nu}$ describes an isotropic surface
sweeped by null string. Taking account of the gauge (6) and relation
(7), the desired Euler--Lagrange motion equations acquire the form
\begin{equation}\label{8}
{\mathaccent "7F x}^M - \left({\g \over \a^\prime
}\right)^2 x^{\prime\prime
M}+\G^M_{PQ}(x)\left[\dot{x}^P\dot{x}^Q-\left({\g \over \a^\prime }
\right)^2 x^{\prime P}x^{\prime Q} \right]=0 \end{equation}
containing the dimensionless parameter
 $\E = \g/\a^\prime $. For the case of closed string Eqs.(8) are completed
by the conditions of periodicity with respect to~$\s $
\begin{equation}\label{9} x^M(\t ,\s =0) = x^M(\t ,\s
=2\pi) \end{equation}
Provided that  $\E \ll 1 $ or, the equivalently 
\begin{equation}\label{10} \dot{x}^M\dot{x}_M \ll
-(x^{\prime M}x^\prime _M) \end{equation} ,
Eqs.(8) may be considered as non--linear ones with the small parameter at
the derivatives with respect to  $\s $. Then we may seek a solution of
Eqs.(8) in the form of a series expansion in terms of the small parameter
$\E $.  In addition, we choose the zero approximation functions $\vp^M(\t )$
independent of $\s $ and satisfying the following equations and constraint
\begin{equation}\label{11} \ddot\vp^M +
 \G^M_{PQ}(\vp)\dot{\vp}^P\dot{\vp}^Q=0 , \end{equation}

\begin{flushright}
\parbox{100mm}{
$(\dot{\vp}^M\dot{\vp}_M) =0$ \hfill (11$^\prime$)}
\end{flushright}
In the zero approximation the constraint (6$^\prime $) is identically
satisfied. Eqs. (11) and       (11$^\prime $)  are equivalent
to the motion equations of a massless particle with the world coordinates
 $\vp^M $ belonging to the geodesic line in the target space--time. It should
be also noted that the constraint (11$^\prime $) is a motion constant.
To derive the equation defining the functions of higher approximations note
that the parameter $\E $ can be eliminated from Eqs.(8) by introducing  the 
new variable   $\xi $ instead of $\s $  using the relations
\begin{equation}\label{12} \xi = \s/\E , ~~~~~~~ \partial_\xi
= \E\partial_\s , ~~~~ \partial_\xi ^2 = \E^2 \partial_\s ^2
\end{equation}
Then the solution of Eqs. (8) may be presented in the form of the series
expansion
  \begin{equation}\label{13} X^M =
 \vp^M(\t)+\E\psi^M(\t,\xi)+\E^2\chi^M(\t,\xi) + ...  \end{equation}
The expansion (13) shows that the effects connected with the string
extension begin to manifest themselves in the the functions of the first 
approximation. To verify this fact it is enough to make the substitution of 
expansion (13) into Eq.(6) for $E$, defining the elastic energy of the
 string tension. After this substitution and the use of the condition
 \begin{equation}\label{14}
 \vp,_\xi^M \equiv {\partial\over \partial \xi }\vp^M =0
 \end{equation}
we will find that the density of the potential energy $E$ (6) acquires the
oscillator-like form
\begin{equation}\label{15}
E=-\g\psi ,^M_\xi G_{MN} \psi ,^N_\xi + ...  \end{equation}
The performed choise of the zero approximation function  $\vp^M $ (14)
essentially simplifies the motion equations and constraints for the
functions of the next approximations. To derive these equations with the
corresponding constraints let us substitute the expansion (13) into Eqs. (8) 
and the constraints (6$^\prime$, 7). Then, taking into account the condition
(14), we obtain the equations for the first approximation
functions $\psi^M$ in the form\footnote{Here and further in the text 
reduced notations are used:
$$\partial_{LK...}G_{MN}\equiv {\partial\over\partial\vp^L}
{\partial\over\partial\vp^K}...G_{MN}(\vp ),~~~~(\vp^M\psi_M)\equiv \vp^M
G_{MN}\psi^N, ... $$ }
\begin{equation}\label{16} D_L^M\psi^L \equiv
\ddot\psi^M-\psi,^M_{\xi\xi}+2[\G_{PQ}^M(\vp )+{1\over 2} \psi^L\partial_L
\G_{PQ}^M] \dot{\vp }^P \dot{\vp }^Q =0~~, \end{equation}

By analogy, retaining the terms of the second order of smallness,the 
equations for the functions of the second order
approximation $\chi^M(\t,\xi)$
\\[4mm]
$D_L^M\chi^L+\G_{PQ}^M\dot{\psi}^P \dot{\psi}^Q
-2\psi^L\partial_L\G_{PQ}^M \dot{\vp}^P\dot{\psi}^Q - {1\over
2}\psi^L\psi^K \partial_{LK}\G_{PQ}^M \dot{\vp}^P\dot{\vp}^Q=0~~,$
\hfill (16$^\prime$),
\\[2mm]
can be obtained.
The equations (16) and (16$^\prime$)  must be completed by the periodicity
conditions with respect to $\xi $. The corresponding expansions of the
constraints (6$^\prime$) and (7) acquire the following form

$$ (\dot{x}^M\cdot
\dot{x}_M)+\left({\g\over \a^\prime}\right)^2 (x^{\prime M} x^\prime_M) =
\varepsilon [2(\dot{\vp}^M\dot\psi_M ) +
\psi^L(\dot{\vp}^M\partial_LG_{MN} \dot{\vp }^N)] +$$

$$
+ \varepsilon^2
[2(\dot{\vp}_M \dot{\chi}^M)  +\chi^L(\dot{\vp}^M \partial_L G_{MN}
\dot{\vp}^N ) +(\dot{\psi}^M \dot{\psi}_M ) +\psi,^M_\xi \psi_{M,\xi}+
$$
\begin{equation}\label{17}
+2\psi^L(\dot{\vp}^M \partial_L G_{MN}\dot{\psi}^N) +{1\over
2}\psi^L \psi^K ( \dot{\vp}^M \partial_{LK} G_{MN}
 \dot{\vp}^N)]+ ... =0 ~~,
\end{equation}

\begin{flushright}

\parbox{130mm}{
$$(\dot{x}^Mx^\prime_M)= \varepsilon (\dot{\vp }^M\psi_{M,\xi})  +
\varepsilon^2[(\dot{\vp}_M\chi
,^M_\xi )+$$
$$+(\dot{\psi}_M \psi ,^M_\xi)+\psi^L(\dot{\vp}_M \partial_L
G_{MN}\psi ,^N_\xi )]+ ...=0 ~~,$$} \hfill (17$^\prime$)
\end{flushright} 

     Note  that in the expansion series (17) and (17$^\prime$) the terms of 
the zero approximation are absent. This fact is explained by the choice of 
the geodesic trajectory (11) and (11$^\prime$) as a zero approximation for the
proper trajectory of the string. The string oscillates  around this geodesic 
trajectory. The approximation in question is applicable when the string 
oscillating energy $\cal{E}$ $\sim$  $1/\sqrt{\a^\prime}$ is small in 
comparison with the mass $M$ $\sim$ $1/\sqrt{\g}$ of the universe where the 
string is living, i.e.  when ${\cal E}/M \sim \sqrt{\g/\a^\prime}<< 1$.
Moreover, note that the oscillator term $\sim x^\prime_Mx^{\prime M}$ in the
primary constraint (7) appears only as a second order perturbation.  As a 
consequence  of this fact the Virasoro terms
     $[(\dot{\psi}^M \dot{\psi}_M)  + ({{\psi^M},}_{\xi}{{\psi_M},}_\xi)]$
in Eq.(17) and correspondently $(\dot{\psi}_M \psi,_\xi^M) $ in Eq.
(17$^\prime$ ), appear starting from the second order approximation.

 The suggested here perturbative approach may be considered as a
 rigorous realization of the heuristic scheme proposed in [2,3,7].
 However, in these papers additional suggestions are made. In particular,
 they use the assumption that the coordinates of string mass center 
 $q^M(\t )$, which play the role of the zero approximation function
 $\vp^M(\t )$, obey the equations of the geodesic line (11) and the constraint
\begin{equation}\label{} (\dot{q}^M \dot{q}_M) =
 (\a^\prime m)^2 , \end{equation}
 where the constant $m$ has the sense of string mass. Our approach is free
 of any assumptions of such kind, because the constraint (11$^\prime$) 
together with other constraints and motion equations, are rigorously derived 
from the unique Euler--Lagrange variational principle for the action (1). The
 constraint  (11$^\prime$) coincides with the constraint (18) only if
  $m=0$.  The latter restriction plays an important role, since it
 requires a more detailed study of the string cosmology effects for which
 the  condition
 $m\not=0$ is essential.

 The author is grateful to E.S.Fradkin, V.P.Frolov, M.Dabrowski,
 S.P.Novikov, G.Veneziano for useful discussions and to S.N.Roshchupkin for
 help. The author wishes to thank N.Sanchez and A.Zichichi for support and
 warm hospitality in the E.Majorana Center in Erice where this paper was
 planned.

 This paper was supported in part by the ISF Grant RY 9200, INTAS Grants
 93--127, 93--633, INTAS and
 Dutch Goverment Grant 94--2317 and the Fund for Fundamental Research of
 CST of Ukraine No 2.3/664.

 \newpage
\vspace{15mm}

 \begin{center} REFERENCES
 \end{center}

\begin{enumerate}
\item G.Veneziano, Status of String Cosmology, Lecture delivered at the
Inter.School "String Gravity and Physics at the Planck Scale", Erice, 8-19
September 1995, Int.Center E.Majorana;\\
M.Gasperini, Metric Perturbation in String Cosmology. Lecture, Ibid.
\item N.Sanchez, String Quantum Gravity. String Theory in Curved Space.
Lecture, Ibid.
\item H.J.de Vega, String Theory in Curved Space. Lecture, Ibid.
\item  N.Sanchez and G,Veneziano, Nucl.Phys.{\bf B333} (1990) 253.
\item M.Gasperini, N.Sanchez and G.Veneziano, Nucl.Phys. {\bf B364} (1991)
265.
\item M.Gasperini,  and G.Veneziano, Phys.Rev. {\bf D50} (1994) 2519.
\item H.J. de Vega and N.Sanchez, Phys.Let. {\bf B197} (1987) 320;\\
Nucl.Phys. {\bf B309} (1988) 577.
\item S.N.Roshchupkin and A.A.Zheltukhin, Class.Quant.Grav. {\bf 12}
(1995) 2519.
\item  A.A.Zheltukhin, Yader.Fiz. {\bf 48}(1988) 375; {\bf 51} (1990) 1504.
\item I.A.Bandos and A.A.Zheltukhin, Fortschr.Phys. {\bf 41} (1993) 619

\end{enumerate}

\end{document}